\documentclass[10pt,journal,compsoc]{IEEEtran}
\ifCLASSOPTIONcompsoc
  \usepackage[nocompress]{cite}
\else
  \usepackage{cite}
\fi
\ifCLASSINFOpdf
\else
\fi
\usepackage{amsmath}
\usepackage{array}
\usepackage{makecell}

\ifCLASSOPTIONcompsoc
  \usepackage[caption=false,font=footnotesize,labelfont=sf,textfont=sf]{subfig}
\else
  \usepackage[caption=false,font=footnotesize]{subfig}
\fi

\usepackage{graphicx}
\usepackage{xcolor}

\usepackage{listings}
\usepackage[hidelinks]{hyperref} % for url links
\hypersetup{
    colorlinks,
    linkcolor={red!50!black},
    citecolor={blue!50!black},
    urlcolor={blue!80!black}
}

\usepackage{xspace} % fixes missing spaces behind custom commands
\newcommand{\textcode}[1]{\textsc{#1}}
\newcommand{\kokkos}{\textcode{Kokkos}\xspace}
\newcommand{\athena}{\textcode{Athena++}\xspace}
\newcommand{\parthenon}{\textcode{Parthenon}\xspace}
\newcommand{\athenaC}{\textcode{Athena}\xspace}
\newcommand{\kathena}{\textcode{K-Athena}\xspace}
\newcommand{\gamer}{\textcode{GAMER}\xspace}
\newcommand{\cuda}{\textcode{CUDA}\xspace}
\newcommand{\openmp}{\textcode{OpenMP}\xspace}
\newcommand{\opencl}{\textcode{OpenCL}\xspace}
\newcommand{\openacc}{\textcode{OpenACC}\xspace}
\newcommand{\mpi}{\textcode{MPI}\xspace}
\newcommand{\mpiopenmp}{\textcode{MPI+OpenMP}\xspace}
\newcommand{\raja}{\textcode{RAJA}\xspace}
\newcommand{\upcpp}{\textcode{UPC++}\xspace}
\newcommand{\charmpp}{\textcode{Charm++}\xspace}
\newcommand{\legion}{\textcode{Legion}\xspace}
\newcommand{\cpp}{\textcode{C++}\xspace}

\newcommand{\gpumembench}{\textcode{GPUMembench}\xspace}
\newcommand{\nvprof}{\textcode{nvprof}\xspace}
\begin{document}
\title{\kathena: a performance portable structured grid finite volume magnetohydrodynamics code }

\author{Philipp~Grete,
        Forrest~W.~Glines,
        and~Brian~W.~O'Shea% <-this % stops a space
\IEEEcompsocitemizethanks{\IEEEcompsocthanksitem P.~Grete, F.~W.~Glines, and B.~.W.~O'Shea 
are with the Department of Physics and Astronomy, Michigan
State University, East Lansing, MI 48824 USA, also with the Department
of Computational Mathematics, Science, and Engineering, Michigan State
University, East Lansing, MI 48824 USA.\protect\\
E-mail: \href{mailto:grete@pa.msu.edu}{grete@pa.msu.edu}
\IEEEcompsocthanksitem B.~W.~O'Shea is also with the National Superconducting Cyclotron 
Laboratory, Michigan State University, East Lansing, MI 48824 USA.
}% <-this % stops an unwanted space
\thanks{Manuscript received \ldots; revised \ldots.}}%


\markboth{Grete \MakeLowercase{\textit{et al.}}: K-Athena 
-- a performance portable structured grid finite volume MHD code}
{Grete \MakeLowercase{\textit{et al.}}: K-Athena
-- a performance portable structured grid finite volume MHD code}



\IEEEtitleabstractindextext{%
\begin{abstract}
Large scale simulations are a key pillar of modern research and require ever-increasing computational resources.
Different novel manycore architectures have emerged in recent years on the way towards the exascale era.
Performance portability is required to prevent repeated non-trivial refactoring of a code for different architectures.
We combine \athena, an existing magnetohydrodynamics (MHD) CPU code,
with \kokkos, a performance portable on-node parallel programming paradigm, into \kathena to allow efficient simulations on multiple architectures using a single codebase.
We  present profiling and scaling results for different platforms including Intel Skylake
CPUs, Intel Xeon Phis, and NVIDIA GPUs.
\kathena achieves $>10^8$ cell-updates/s on a single V100 GPU
for second-order double precision MHD calculations, and a speedup of 30 on up to 24,576 GPUs
on Summit (compared to 172,032 CPU cores), reaching  $1.94\times10^{12}$ total
cell-updates/s at $76\%$ parallel efficiency.
Using a roofline analysis we demonstrate that the overall performance is currently limited 
by DRAM bandwidth and calculate a performance portability metric of 62.8\%.
Finally, we present the implementation strategies used and the challenges
encountered in maximizing performance.
This will provide other research groups with a straightforward approach to prepare
their own codes for the exascale era.
\kathena is available at \url{https://gitlab.com/pgrete/kathena}.
\end{abstract}

}

\maketitle

\IEEEdisplaynontitleabstractindextext

% \fi
\IEEEpeerreviewmaketitle

\IEEEraisesectionheading{\section{Introduction}\label{sec:introduction}}

\IEEEPARstart{T}{he} era of exascale computing is approaching.
Different projects around the globe are working on the first exascale
supercomputers, i.e., supercomputers capable of conducting $10^{18}$
 floating point operations per second.
This includes, for example, the Exascale Computing Initiative working
with Intel and Cray on Aurora as the first exascale computer in the US in 2021,
the EuroHPC collaboration working on building two exascale systems in Europe by
2022/2023, Fujitsu and RIKEN in Japan working on the Post-K machine to launch in 
2021/2022, and China who target 2020 for their first exascale machine.
While the exact architectural details of these machines are not announced yet and/or are still
under active development, the overall trend in recent years has been manycore 
architectures.
Here, manycore refers to an increasing number of (potentially simpler) cores on a single
compute node and includes CPUs 
(e.g., Intel's Xeon Scalable Processor family or  AMD's Epyc family),
accelerators (e.g., the now discontinued Intel Xeon Phi line), and GPUs for general
purpose computing.
\mpiopenmp has been the prevailing parallel programming paradigm in many areas 
of high performance computing for roughly two decades.
It is questionable, however, whether this generic approach will be capable
of making efficient use of available hardware features such as parallel threads and
vectorization across different manycore architectures and between nodes.

%\IEEEpubidadjcol
In addition to extensions of the \mpi standard such as shared-memory parallelism, 
several approaches in addition to \mpiopenmp exist and are being actively developed to address 
either on-node, inter-node, or both types of parallelism.
These include, for example, partitioned global address space (PGAS) programming models
such as \upcpp \cite{upcpp}, or parallel programming frameworks such as
\charmpp or \legion, which are based on  message-driven migratable
objects \cite{charmpp,legion}.

Our main goal is a performance portable version of the existing \mpiopenmp 
finite volume (general relativity)
magnetohydrodynamics (MHD) code \athena \cite{White2016,Stone2019}. 
This goal includes enabling GPU-accelerated simulations while
maintaining CPU performance using a single code base.
More generally, performance portability refers to achieving consistent levels
of performance across heterogeneous platforms using as little architecture-dependent
code as possible.
Given the uncertainties in future architecures (and the broad availability of different
architecture already today) performance portability is an active field of research
in many areas \cite{perfportbook,Bennett2015}.
This includes (but is not limited to) idealized benchmarks and miniapps 
\cite{Heroux2009, Martineau2017, Deakin2018, Hammond2019},
algorithm libraries \cite{Heroux2012},
structured mesh codes \cite{Holmen2019}, or particle in cell codes \cite{Artigues2019}.

In order to keep the code changes minimal, and given the \mpiopenmp basis of \athena, we
decided to keep \mpi for inter-node parallelism and focus on on-node performance
portability.
For on-node performance portability several libraries and programming language extensions
exist.
With version 4.5 \openmp \cite{openmp} has been extended to support offloading to devices
such as GPUs, but support and maturity is still highly compiler and 
architecture dependent.
This similarly applies to \openacc, which
has been designed from the beginning to target heterogeneous platforms.
While these two directives-based programming models are generally less intrusive 
with respect to the code base, they only expose a limited fraction of various platform-specific features.
\opencl \cite{opencl} is much more flexible and allows fine grained control over
hardware features (e.g., threads), but this, on the other hand, 
adds substantial complexity to the code.
\kokkos \cite{Edwards2014} and \raja \cite{raja} try to combine the strength of 
flexibility with ease of use by providing abstractions in the form of \cpp templates.
Both \kokkos and \raja focus on abstractions of parallel regions in the code, and
\kokkos additionally provides abstractions of the memory hierarchy.
At compile time the templates are translated to different (native) backends, e.g., 
\openmp on CPUs or \cuda on NVIDIA GPUs.
A more detailed description of these different approaches including benchmarking
in more idealized setups can be found in, e.g., \cite{Martineau2017,Deakin2018}.

We chose \kokkos for the refactoring of \athena for several reasons.
\kokkos offers the highest level of abstraction without forcing the developer to use it
by setting reasonable implicit platform defaults.
Moreover, the \kokkos core developer team actively works on integrating the 
programming model into the \cpp standard.
New, upcoming features, e.g., in \openmp, will replace manual implementations in the 
\kokkos \openmp backend over time.
\kokkos is already used in several large projects to achieve performance portability, e.g.,
the scientific software building block collection Trilinos \cite{Heroux2005} or 
the computational framework for simulating chemical and physical reactions Uintah 
\cite{Holmen2017}.
In addition, \kokkos is part of the DOE's Exascale Computing Project and we thus expect 
a backend for Aurora's new Intel Xe architecture when the system launches.
Finally, the \kokkos community, including core developers and users, is very active and
supportive with respect to handling issues, questions and offering workshops.

The resulting \kathena code successfully achieves performance portability across
CPUs (Intel, AMD, and IBM), Intel Xeon Phis, and NVIDIA GPUs.
We demonstrate weak scaling at 76\% parallel efficiency on 24,576 GPUs on OLCF's
Summit, reaching $1.94\times10^{12}$ total cell-updates/s for a double
precision MHD calculation.
Moreover, we calculate a performance portability metric of 62.8\% across Xeon Phis,
6 CPU generations, and 3 GPU generations.
We make the code available as an open source project\footnote{
\kathena's project repository is located at \url{https://gitlab.com/pgrete/kathena}.}.

The paper is organized as follows. 
In Section~\ref{sec:method} we introduce \kokkos, \athena, and the changes made 
and approach chosen in creating \kathena.
In Section~\ref{sec:results} we present profiling, scaling and roofline
analysis results.
Finally, we discuss current limitations and future enhancements in 
Sec.~\ref{sec:limit-and-enhance} and make concluding remarks in Sec.~\ref{sec:conclusions}.

\section{Method}
\label{sec:method}
\subsection{\kokkos}
\label{sec:kokkos}
\kokkos is an open source\footnote{
  See \url{https://github.com/kokkos} for the library itself, associated tools, 
  tutorial and a wiki.}
\texttt{C++} performance portability programming model 
\cite{Edwards2014}.
It is implemented as a template library and offers abstractions for parallel execution
of code and data management.
The core of the programming model consists of six abstractions.

First, \textit{execution spaces} define where code is executed. This includes, for example,
\openmp on CPUs or Intel Xeon Phis, \cuda on NVIDIA GPUs, or \texttt{ROCm} on AMD GPUs 
(which is currently experimental).
Second, \textit{execution patterns} are parallel patterns, e.g. \texttt{parallel\_for} or 
\texttt{parallel\_reduce}, are the building blocks of any application that uses
\kokkos.
These parallel regions are often also referred to as kernels as they can be dispatched for 
execution on execution spaces (such as GPUs).
Third, \textit{execution policies} determine how an execution pattern is executed.
There exist simple range policies that only specify the indices of
the parallel pattern and the order of iteration (i.e., the fastest changing index for
multidimensional arrays).
More complicated policies, such as team policies, can be used for more fine-grained
control over individual threads and nested parallelism.
Fourth, \textit{memory spaces} specify where data is located, e.g., in host/system
memory or in device space such as GPU memory.
Fifth, the \textit{memory layout} determines the logical mapping of multidimensional
indices to actual memory location, cf., \texttt{C} family row-major order versus
\texttt{Fortran} column-major order.
Sixth, \textit{memory traits} can be assigned to data and specify how data is accessed, e.g.,
atomic access, random access, or streaming access.

These six abstractions offer substantial flexibility in fine-tuning
application, but the application developer is not always required to specify all details.
In general, architecture-dependent defaults are set at compile time based on the 
information on devices and architecture provided.
For example, if \cuda is defined as the default execution space at compile time, all
\texttt{Kokkos::View}s, which are the fundamental multidimensional array structure,
will be allocated in GPU memory.
Moreover, the memory layout is set to column-major so that consecutive threads in the 
same warp access consecutive entries in memory.

\subsection{\athena}
\label{sec:athena}
\athena is a radiation general relativistic magnetohydrodynamics (GRMHD) code focusing on astrophysical 
applications~\cite{White2016,Stone2019}.
It is a rewrite in modern \cpp of the widely used \athenaC \texttt{C} version 
\cite{Stone2008}.
\athena offers a wide variety of compressible hydro- and magnetohydrodynamics solvers 
including support for special and relativistic (M)HD, 
flexible geometries (Cartesian, cylindrical, or spherical), and mixed parallelization
with \openmp and \mpi.
Apart from the overall feature set, the main reasons we chose \athena are 
a) its excellent performance on CPUs and KNLs due to a focus on vectorization
in the code design, 
b) a generally well written and documented code base in modern 
\cpp, 
c) point releases are publicly 
available that contain many (but not all) features\footnote{Our code
  changes are based on the public version, \athena 1.1.1, see
\url{https://github.com/PrincetonUniversity/athena-public-version}},
and
d) a flexible task-based execution model that allows for a high degree of modularity.

\athena's parallelization strategy evolves around so-called \texttt{meshblocks}.
The entire simulation grid is divided into smaller \texttt{meshblocks} that are distributed
among \mpi processes and/or \openmp threads.
Each \mpi processes (or \openmp thread) owns one or more \texttt{meshblocks} that can
be updated independently after boundary information have been communicated.
If hybrid parallelization is used, each \mpi process runs one or more \openmp threads that
each are assigned one or more \texttt{meshblock}.
This design choice is often referred to as coarse-grained parallelization as threads are used
at a block (here \texttt{meshblock}) level and not over loop indices.
In general, \athena uses persistent \mpi communication handles 
in combination with one-sided \mpi calls to realize asynchronous 
communication.
Moreover, each thread makes its own \mpi calls to exchange boundary information.
As a result, using more than one thread per \mpi process may increase overall on-node
performance due to hyperthreading but also increases both the number of \mpi messages sent and
the total amount of data sent.
The latter may result in overall worse parallel performance and efficiency, as
demonstrated in Sec~\ref{sec:weak-scaling}.

\begin{lstlisting}[language=C++,label=lst:simd-for,caption={
Example triple \lstinline!for! loop for a typical
operation in a finite volume method on a structured mesh such as in a
code like \athena, where \lstinline!ks!, \lstinline!ke!, \lstinline!js!,
\lstinline!je!, \lstinline!is!, and \lstinline!ie! are loop bounds and
\lstinline!u! is an \texttt{athena\_array} object of, for example,
an MHD variable.
}]

for( int k = ks; k < ke; k++){
  for( int j = js; j < je; j++){
    #pragma omp simd
    for( int i = is; i < ie; i++){
      /* Loop Body */
      u(k,j,i) = ...
}}}
\end{lstlisting} 
Given the coarse-grained \openmp approach over \texttt{meshblocks} the prevalent 
structures in the code base are triple (or quadruple) nested \texttt{for} loops
that iterate over the content of each \texttt{meshblock} (and variables in the quadruple
case).
A prototypical nested loop is illustrated in Listing~\ref{lst:simd-for}.
Generally, all loops (or kernels) in \athena have been written so that \openmp 
\texttt{simd pragmas} are used for the innermost loop.
This helps the compiler in trying to automatically vectorize the loops resulting
in a more performant application.

\subsection{\kathena = \kokkos + \athena}
\label{sec:kathena-loops}

In order to combine \athena and \kokkos, four major changes in the
code base were required: 1) making \texttt{Kokkos::View}s the fundamental
data structure, 2) converting nested \texttt{for} loop structures to kernels,
3) converting ``support'' functions, such as the equation of state, to inline functions, and
4) converting communication buffer filling functions into kernels.

First, \texttt{View}s are the \kokkos' abstraction of multidimensional arrays.
Thus, the multidimensional arrays originally used in \athena, e.g., the MHD variables
for each \texttt{meshblock}, need to be converted to \texttt{View}s so that these
arrays can transparently be allocated in arbitrary memory spaces such as device (e.g., GPU)
memory or system memory.
\athena already implemented an abstract \texttt{athena\_array} class for all
multidimensional arrays with an interface similar to the interface of a \texttt{View}.
Therefore, we only had to add \texttt{View} objects as member variables and 
to modify the functions of \texttt{athena\_array}s to transparently use 
functions of those member \texttt{View}s.
This included using \texttt{View} constructors to allocate memory, using
\texttt{Kokkos::deep\_copy} or \texttt{Kokkos::subview} for copy constructors and 
shallow slices, and creating public member functions to access the \texttt{View}s.
The latter is required in order to properly access the data from within compute kernels.

Second, all nested \texttt{for} loop structures (see Listing~\ref{lst:simd-for}
need to be converted to so-called kernels, i.e., parallel region that can be dispatched
for execution by an execution space.
As described in Sec.~\ref{sec:kokkos} multiple execution policies are possible,
such as a multidimensional range policy (see Listing~\ref{lst:MDRange}),
a one dimensional policy with manual index mapping (see Listing~\ref{lst:1DRange}),
or a team policy
that allows for more fine-grained control and nested parallelism 
(see Listing~\ref{lst:TeamPolicy}).

\begin{lstlisting}[language=C++,label=lst:MDRange,caption={
Example \lstinline!for! loop using \kokkos. The loop body is
reformulated into a lambda function and passed into
\lstinline!Kokkos::parallel_for!  to execute on the target architecture.
The class \lstinline!Kokkos::MDRangePolicy! specifies the loop bounds.
The  array \lstinline!u! is now a \lstinline!Kokkos::View!, a
\kokkos building block that allows transparent access to CPU and GPU memory.
The loop body, i.e., the majority of the code, remains mostly unchanged.
}]

parallel_for( MDRangePolicy<Rank<3>> 
    ({ks,js,is},{ke,je,ie}),
  KOKKOS_LAMBDA(int k, int j, int i){
    /* Loop Body */
    u(k,j,i) = ...
});
\end{lstlisting} 

\begin{lstlisting}[language=C++,label=lst:1DRange,caption={
Same as Listing~\ref{lst:MDRange} but using a one dimensional
\lstinline!Kokkos::RangePolicy! (implicit through default template
parameter) with explicit index calculation.
}]

int nk = ke-ks, nj = je-js, ni = ie-is;
parallel_for(nk*nj*ni, 
  KOKKOS_LAMBDA(int idx){
    int k = idx / (nj*ni);
    int j = (idx - k*(nj*ni) / ni;
    int i = idx - k*(nj*ni) - j*ni;
    /* Loop Body */
    u(k,j,i) = ...
});
\end{lstlisting} 

\begin{lstlisting}[language=C++,label=lst:TeamPolicy,caption={
Another approach using \kokkos' nested team-based parallelism
through the \lstinline!Kokkos::TeamThreadRange! and
\lstinline!Kokkos::ThreadVectorRange! classes. This interface is
closer to the underlying parallelism used by the backend such as \cuda
blocks on GPUs and SIMD vectors on CPUs.
}]

parallel_for(team_policy(nk, AUTO),
  KOKKOS_LAMBDA(member_type thread) {
    const int k = thread.league_rank() + ks;
    parallel_for(
      TeamThreadRange<>(thread,js,je,
        [&] (const int j) {
        parallel_for(
          ThreadVectorRange<>(thread,is,ie,
          [=] (const int i) {
            /* Loop Body */
            u(k,j,i) = ...
});});});
\end{lstlisting}
Generally, the loop body remained mostly unchanged.
Given that it is not a priori clear what kind of execution policy yields the best performance
for a given implementation of an algorithm, we decided to implement a flexible loop 
  macro\footnote{Note, that in newer versions of the code we replaced the macro with
  a template.}.
That macro allows us to easily change the execution policy for
performance tests -- see
profiling results in Sec.~\ref{sec:single-scaling} 
and discussion in Sec.~\ref{sec:limit-and-enhance},
and this intermediate abstraction is similar to the approach
chosen in other projects \cite{Holmen2019}.

Third, all functions that are called within a kernel need to be converted into inline
functions (here, more specifically using the \texttt{KOKKOS\_INLINE\_FUNCTION} macro).
This is required because if the kernels are executed on a device such as a GPU, the
function need to be compiled for the device (e.g., with a \texttt{\_\_device\_\_} attribute when compiling with \cuda).
In \athena, this primarily concerned functions such as the equation of state and
coordinate system-related functions.

Fourth, \athena uses persistent communication buffers (and \mpi handles) to exchange data
between processes.
Originally, these buffers resided in the system memory and were filled directly from
arrays residing in the system memory.
In the case where a device (such as a GPU) is used as the primary execution space and the arrays should
remain on the device to reduce data transfers, the buffer filling functions need to be 
converted too.
Thus, we changed all buffers to be \texttt{View}s and converted the buffer filling 
functions into kernels that can be executed on any execution space.
In addition, this allows for \cuda-aware \mpi -- GPU buffers to be directly
copied between the memories of GPUs (both on the same node and on different nodes)
without an implicit or explicit copy of the data to system memory.

In general, the first three changes above are required in refactoring any legacy code
to make use of \kokkos.
We note that the original \athena design made it mostly straightforward to
implement those changes, e.g., because of the existence of an abstract array class and
the prevailing tightly nested loops already optimized for vectorized instructions.
More broadly, we expect that structured grid fluid codes will require similar changes
and that other algorithms and application may require more subtle refactoring in
order to achieve good performance.
The fourth change was required more specifically for \athena due to the 
existing \mpi communication patterns.

Finally, for the purpose of the initial proof-of-concept, we only refactored the parts
required for running hydrodynamic and magnetohydrodynamic simulations on 
static and adaptive Cartesian meshes.
Running special and general relativistic simulations on
spherical or cylindrical coordinates is currently not supported.
However, the changes required to allow for these kind of simulations are straightforward
and we encourage and support contributions to re-enable this functionality.

Throughout the development process, we continuously measured the code performance
in detail using so-called \kokkos profiling regions as well as the automated
profiling of all \kokkos kernels.
Moreover, we employed automated regression testing using GitLab's continuous 
integration features and included specific tests to address changes related to \kokkos
(such as running on different architectures and testing different loop patterns).

\section{Results}
\label{sec:results}

If not noted otherwise, all results in this section have been obtained using
a double precision, shock-capturing, unsplit, adiabatic MHD solver consisting 
of Van Leer integration,
piecewise linear reconstruction, Roe Riemann solver, and constrained transport for the 
integration of the induction equation (see, e.g., \cite{Stone2009} for more details).
The test problem is a linear fast magnetosonic wave on a static, structured,
three-dimensional grid.
In GPU runs there is no explicit data transfer between system and GPU memory except
during problem initialization, i.e., the exchange of ghost cells is handled either
by direct copies between buffers in GPU memory on the same GPU or between buffers in
GPU memory on different GPUs using \cuda-aware \mpi.
Similarly, there is also no implicit data transfer as unified memory was not used.
Generally, we used the Intel compilers on Intel platforms, and \texttt{gcc}  and 
\texttt{nvcc} on other platforms as we found that (recent) Intel compilers are
more effective in automatic vectorization than (recent) \texttt{gcc} compilers.
We used the identical software environment and compiler flags
for both \kathena and \athena where possible.
Details are listed in Table~\ref{tab:compiler}.
We used \athena version 1.1.1 (commit \texttt{4d0e425}) and \kathena commit 
\texttt{73fec12d} for the scaling tests.
Additional information on how to run \kathena on different machines can be found in
the code's documentation.
\begin{table*}[!t]
\renewcommand{\arraystretch}{1.3}
\caption{Software Environment and Compiler Flags Used In Scaling Tests.}
\label{tab:compiler}
\centering
\begin{tabular}{l c p{8cm} c}
\hline
\bfseries Machine & \bfseries Compiler & \bfseries Compiler flags & \bfseries \mpi version\\
\hline
Summit GPU & GCC 6.4.0 \& Cuda 9.2.148 & 
\texttt{-O3 -std=c++11 -fopenmp -Xcudafe --diag\_suppress=esa\_on\_defaulted\_function\_ignored -expt-extended-lambda -arch=sm\_70 -Xcompiler}
& Spectrum MPI 10.2.0.11 \\
Summit CPU & GCC 8.1.1  & 
\texttt{-O3 -std=c++11 -fopenmp-simd -fwhole-program -flto -ffast-math -fprefetch-loop-arrays -fopenmp -mcpu=power9 -mtune=power9}
& Spectrum MPI 10.2.0.11 \\
Titan GPU & GCC 6.3.0 \& Cuda 9.1.85 &
\texttt{-O3 -std=c++11 -fopenmp -Xcudafe --diag\_suppress=esa\_on\_defaulted\_function\_ignored -expt-extended-lambda -arch=sm\_35 -Xcompiler}
& Cray MPICH 7.6.3 \\
Titan CPU & GCC 6.3.0 & 
\texttt{-O3 -std=c++11 -fopenmp}
& Cray MPICH 7.6.3 \\
Theta & ICC 18.0.0 &
\texttt{-O3 -std=c++11 -ipo -xMIC-AVX512 -inline-forceinline -qopenmp-simd -qopenmp}
& Cray MPICH 7.7.3\\
Electra & ICC 18.0.3 &
\texttt{-O3 -std=c++11 -ipo -inline-forceinline -qopenmp-simd -qopt-prefetch=4 -qopenmp -xCORE-AVX512}
& HPE MPT 2.17\\
\hline
\end{tabular}
\end{table*}

\subsection{Profiling}
\label{sec:profiling}
\begin{figure*}[htbp]
\centering
\includegraphics[width=\textwidth]{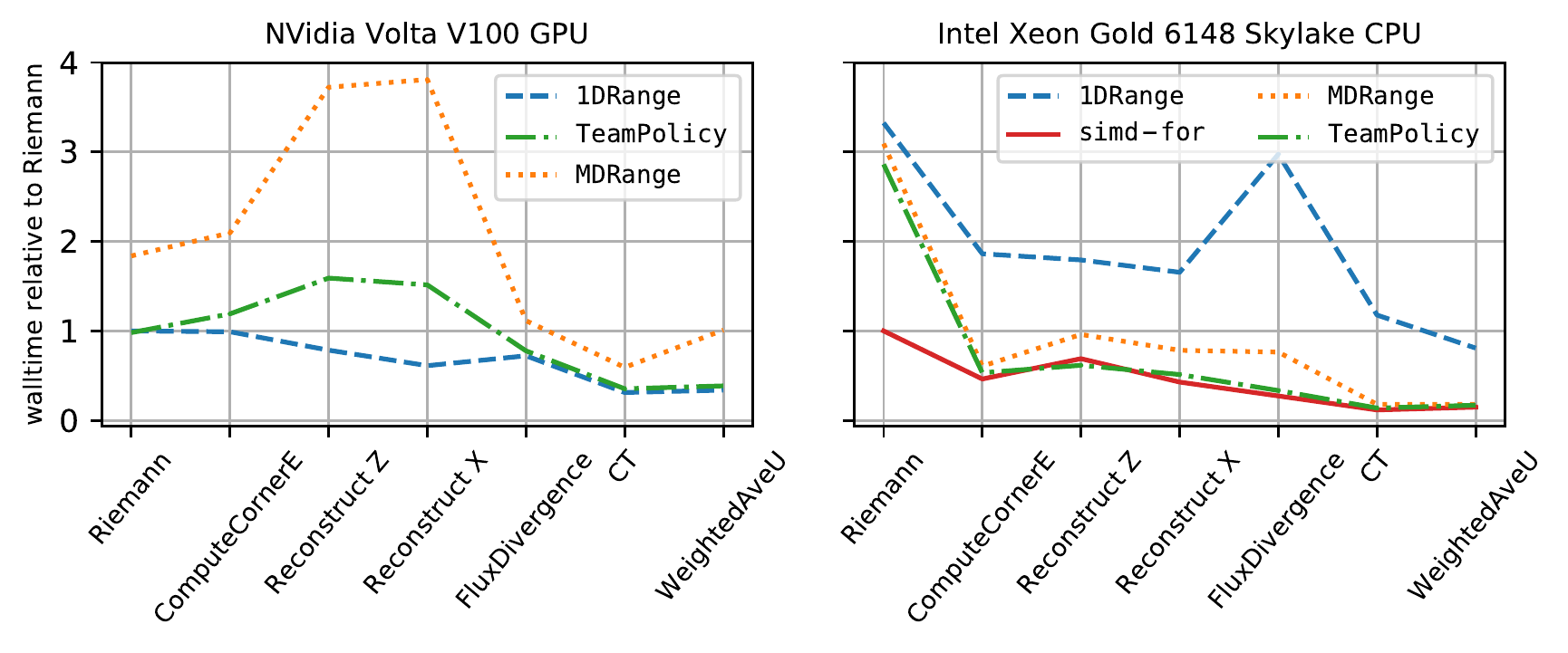}
\caption{Profiling results on a GPU (left) and CPU (right) for selected regions (x-axis)
within the main loop of an MHD timestep 
using the algorithm described in Sec.~\ref{sec:results}.
The different lines correspond to different loop structures, see Sec.~\ref{sec:kathena-loops}
and the timings are normalized to the fastest Riemann region in each panel.
}
\label{fig:regions}
\end{figure*}

In order to evaluate the effect on performance of the different loop structures presented in
Sec.~\ref{sec:kathena-loops} we compare the timings of different regions within the main
loop of the code.
The results using both an NVIDIA V100 GPU and an Intel Skylake CPU for a selection of the 
computationally most expensive regions are shown in Fig.~\ref{fig:regions}.
The \texttt{1DRange} loop structure refers to a one dimensional range policy over a single
index that is explicitly unpacked to the multidimensional indices in the code
(cf.~Listing~\ref{lst:1DRange}).
While this \texttt{1DRange} is the fastest loop structure for all regions on the GPU, it
is the slowest for all regions on the CPU.
According to the compiler report this particular one dimensional mapping prevents 
automated vectorization optimizations.
All other loop structures tested, i.e., \texttt{simd-for}
(cf.~Listing~\ref{lst:simd-for}),
\texttt{MDRange} (cf.~Listing~\ref{lst:MDRange}), 
and \texttt{TeamPolicy} (cf.~Listing~\ref{lst:TeamPolicy}) 
logically separate the nested loops and, thus, make it easier 
for the compiler to automatically vectorize the innermost loop.
This also explains why the results for \texttt{simd-for}, \texttt{MDRange}, and 
\texttt{TeamPolicy} are very close to each other for all regions except the Riemann solver.
The Riemann solver is the most complex kernel in the chosen setup so that the
compiler is not automatically vectorizing this loop despite the \verb=#pragma ivdep=
in \kokkos' \texttt{MDRange} and \texttt{TeamPolicy}.
Only the more aggressive explicit \verb=#pragma omp simd= results in a vectorized loop.
The aggregate performance differences (all kernels of a cycle combined) to the fastest 
\texttt{simd-for} pattern are 0.78 (\texttt{TeamPolicy}), 0.71 (\texttt{MDRange}), and
0.51 (\texttt{1DRange}).

On the GPU, \texttt{MDRange} is the slowest loop structure, being several times (2x-4x)
slower than the \texttt{1DRange} across all regions.
\texttt{TeamPolicy} is on par with \texttt{1DRange} for half of the regions shown.
Here, the aggregate performance differences to the fastest \texttt{1DRange} pattern
are 0.75 (\texttt{TeamPolicy}) and 0.078 (\texttt{MDRange}). 
As discussed in more detail in Sec.~\ref{sec:limit-and-enhance}, we expected these
non-optimized raw loop structures to not cause any major differences in performance.

The results shown here for V100 GPUs and Skylake CPUs equally apply to other
GPU generations and other CPUs (and Xeon Phis), respectively.
For all tests conducted in the following, we use the loop structure with the highest
performance on each architecture, i.e., \texttt{1DRange} on GPUs and \texttt{simd-for}
on CPUs and Xeon Phis.

\subsection{Performance portability}
\label{sec:performance-portability}

Our main objective for writing \kathena is an MHD code that runs efficiently on
any current supercomputer and possibly any future machines.  A code
that runs efficiently on more architectures is said to be performance portable.
Determining what is meant by ``efficient code'' can be vague, especially when
comparing performance across different architectures. The memory space sizes,
bandwidths, instruction sets, and arrangement of cores on different architectures
can all affect how efficiently a code can utilize the hardware.  

In order to make fair comparisons of \kathena's performance across different
machines (see Sec.~\ref{sec:arch}), we used the roofline model \cite{williams_roofline_2009}, described
in Sec. \ref{sec:roofline-models}, to compute on several architectures the
architectural efficiency of \kathena, or the fraction of the performance
achieved compared to the theoretical performance as limited hardware.  We then
used the architectural efficiencies to compute the performance portability
metric from \cite{Pennycook2019}, described in Sec.
\ref{sec:performance-portability-metric}, to quantify the
performance portability of \kathena.

\subsubsection{Overview of architectures used}
\label{sec:arch}
\begin{table*}[!htbp]
\renewcommand{\arraystretch}{1.3}
\caption{
  Technical specifications for devices used in the performance portability
  metric. Cache size and core counts for CPUs specify the aggregate sizes and
  counts for a two-socket node while numbers for GPUs show the aggregate for a
  single device. For the Tesla K80, the cache size and core count is for just
  one of the two GK210 chips in the GPU. For DRAM bandwidth we use the
  empirically measured bandwidth of the DRAM on CPUs and the global memory on
  GPUs. Data for Intel devices comes from \cite{intel_optimization_manual} and
  data for NVIDIA devices comes from
  \cite{kepler_whitepaper,pascal_whitepaper,volta_whitepaper,jia_dissecting_2018}.
  }
\label{tab:device_specifications}
\centering
  \scriptsize
\begin{tabular}{ l | p{1.10cm} p{1.10cm} p{1.10cm} p{1.10cm} p{1.10cm} p{1.10cm} p{1.10cm} p{1.10cm} p{1.10cm} p{1.10cm}  }
\bfseries Manufacturer & Intel & Intel & Intel & Intel & Intel & Intel & Intel & NVIDIA & NVIDIA & NVIDIA\\
\bfseries Family & Xeon E5 & Xeon E5 & Xeon E5 & Xeon E5 & Xeon Gold & Xeon Gold & Xeon Phi & Tesla & Tesla & Tesla\\
\bfseries Microarchitecture & Sandy Bridge & Ivy Bridge & Haswell & Broadwell & Skylake & Cascade Lake & Knights Landing & Kepler & Pascal & Volta\\
\bfseries Model & 2670 & 2680v2 & 2680v3 & 2680v4 & 6148 & 6248 & 7250 & K80 & P100 & V100\\
\bfseries Instruction Set & AVX & AVX & AVX2 & AVX2 & AVX512 & AVX512 & AVX512 &  &  & \\
\bfseries CUDA Capability &  &  &  &  &  &  &  & 3.7 & 6.0 & 7.0\\
\bfseries Clock Rate (GHz)& 2.6 & 2.8 & 2.5 & 2.4 & 2.4 & 2.5 & 1.4 & 0.562 & 1.328 & 1.29\\
\bfseries Num. Cores & 16 & 20 & 24 & 28 & 40 & 40 & 68 & 832 & 1792 & 2560\\
\bfseries Max L1 Data Cache (KB)& 512 & 640 & 768 & 896 & 1280 & 1280 & 2176 & 1456 & 1344 & 10240\\
\bfseries Total L2 Data Cache (KB)& 4096 & 2560 & 5120 & 7168 & 40000 & 40000 & 34000 & 1536 & 4096 & 6144\\
\bfseries Total L3 Data Cache (MB)& 40 & 50 & 60 & 70 & 55 & 55 &  &  &  & \\
\bfseries DRAM Bandwidth (GB/s)& 97.9 & 121 & 139 & 147 & 246 & 247 & 494 & 195 & 521 & 782
\end{tabular}
\end{table*}
In total, we created roofline models for six Intel CPUs, Intel Xeon Phis, and three
NVIDIA GPUs.
The CPU models roughly follow Intel's tick-tock production model and, thus, span
pairs of three different instructions sets (AVX, AVX2, and AVX512) with one CPU introducing
a new instruction set and the other an increase in cores and/or clock rate
with the same instruction set.
The Intel Xeon Phi (Knights Landing) also supports AVX512 instructions and differs
from the CPUs at the highest level by an increased core count, lower clock rate, and access
to MCDRAM.
The three different NVIDIA GPUs span three different microarchitectures 
(Kepler, Pascal, and Volta), which also translates to an increased core count in the
GPUs used.
L1 data caches are also implemented differently across the three
microarchitectures.  
On Kelper and Volta GPUs, the L1 cache is physically in the same memory
device as CUDA "shared" memory while on Pascal GPUs the L1 cache is combined
with texture memory
\cite{kepler_whitepaper,pascal_whitepaper,volta_whitepaper}. 
Load throughput to L1 cache on Pascal GPUs achieves lower bytes/cycle compared
to Kelper and Volta GPUs \cite{jia_dissecting_2018},  which led to K-Athena
maintaining a higher fraction of peak L1 bandwidth.
An comparative overview of the technical specifications for all architectures is
given in Table~\ref{tab:device_specifications}.

\subsubsection{Roofline model}
\label{sec:roofline-models}
\begin{figure}[!t]
  \centering
  \subfloat[Cascade Lake CPU Roofline]{
    \centering
    \includegraphics[width=3.5in]{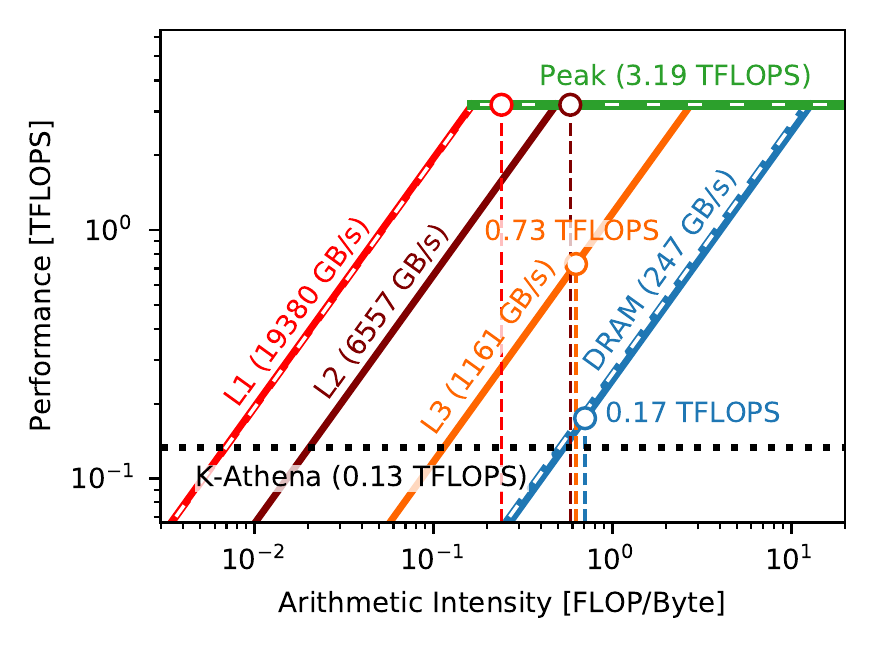}
    \label{fig:cpu-roofline}
  }

  \subfloat[Tesla V100 GPU Roofline]{
    \centering
    \includegraphics[width=3.5in]{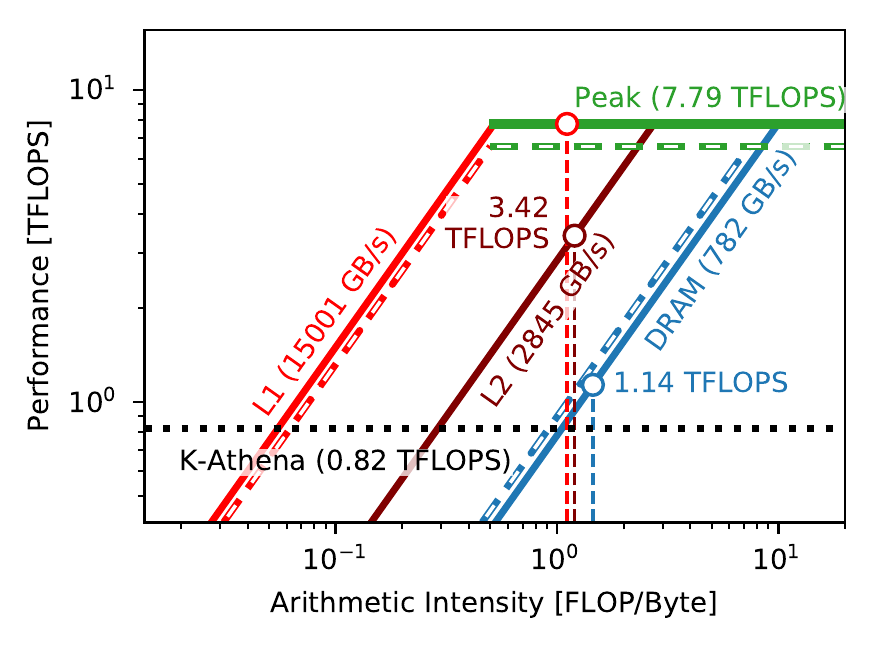}
    \label{fig:gpu-roofline}
  }

  \caption{ Roofline models of a 2 socket Intel Xeon Gold 6248 "Cascade Lake" CPU
  node on NASA's Aitken (\ref{fig:cpu-roofline}) and a single NVIDIA Tesla
  V100 "Volta" GPU on MSU HPCC (\ref{fig:gpu-roofline}). Theoretical L1 and
  DRAM bandwidths and theoretical peak throughputs according to manufacturer
  specifications are shown in dashed line. for For both cases shown here and
  all other architectures we tested, DRAM bandwidth (or MCDRAM bandwidth for
  KNLs) is the limiting bandwidth for \kathena's performance. 
  }
  \label{fig:rooflines}
\end{figure}

The roofline model is a graphical tool to demonstrate the theoretical peak performance of
an application on an architecture by condensing the performance limits imposed
by the bandwidth of each memory space and peak throughput of the device into a
single plot.  In a roofline model plot, peak throughputs and bandwidths
of the hardware are plotted on a log Performance [FLOPS] versus log arithmetic
intensity [FLOP/B] axis so that throughputs are horizontal lines and
bandwidths as $P\propto I$ lines (since bandwidth-limited $P = B \times I$),
where $P~\text{[FLOPS]}$ is performance\footnote{In this work we consider double precision
throughput and count FMA instructions as two FLOP on architectures that support
it.}, $I ~\text{[FLOP/B]}$ is arithmetic intensity (the operations executed per
byte read and written), and $B ~\text{[B/s]}$ is the bandwidth.
The arithmetic intensities of each memory space for a specific
application appear as vertical lines, extending up where the bandwidth of the memory space limits
performance.

The maximum theoretical performance of an application is limited by the
bandwidth and throughput ceilings displayed in the roofline model. For the
given device and application, the maximum obtainable performance in FLOPS is
limited by
\begin{align}
  P_{\text{max}}(a,p,i) \leq 
    \min_{m \in M} \left \{ 
    \min \left [ \right . \right . &
       T_{\text{Peak}}(i),   \\ \notag
        & \left . \left .
       B(i,m) \times I(a,p,i,m) 
       \right ] \right \},
\end{align}
where $P_{\text{max}}(a,p,i) \text{[FLOPS]}$ is the maximum possible FLOPS
obtainable by application $a$ solving problem $p$ on architectural platform
$i$, $T_{\text{Peak}}(i) \text{[FLOPS]}$ is the peak throughput on the
platform,  $M$ is all the memory spaces on the device (L1 cache, L2 cache,
DRAM, etc.), and $I(a,p,i,m) \text{[FLOP/B]}$ is the arithmetic intensity the
application solving the problem on the memory space $m$, or the number of FLOP
executed per number of bytes written and read to and from $m$.  We can also
mark the actual performance of application with a horizontal dashed line,
indicating the actual average FLOPS achieved.  Figures~\ref{fig:cpu-roofline}
and~\ref{fig:gpu-roofline} show roofline models of \kathena solving a $256^3$
linear wave on an Intel Cascade Lake CPU node on NASA's Aitken and a single
NVIDIA Volta V100 GPU on MSU's HPCC.

Using the roofline model, we can quantify the architectural efficiency of the
\kathena, or the fraction of performance achieved compared to the theoretical
maximum performance of the algorithm as limited by bandwidth. In this work, we
further distinguish multiple architectural efficiencies per platform as limited
by the bandwidth of different memory spaces. The architectural efficiency
$e(a,p,i,m)$ of the application $a$ solving the problem $p$ on platform $i$ as
limited by the bandwidth of the memory space $m$ on platform $i$ is 
\begin{equation} 
  e(a,p,i)
  = \frac{\varepsilon(a,p,i)}{\min \left ( T_{\text{Peak}}(i), B(i,m) \times
I(a,p,i,m) \right ) } 
\end{equation} 
where $\varepsilon(a,p,i)$ is the
achieved performance of the application $a$ for solving the problem $p$ on the
platform $i$, $B(i,m)$ is the peak DRAM bandwidth on the platform, and
$I(a,p,i,m)$ is the arithmetic intensity of the  for solving the problem on
that platform.  For example, on Summit's Volta V100s, \kathena achieves $0.82$
TFLOPS while the DRAM bandwidth limits performance to $1.13$ TFLOPS, giving to a
$72.5\%$ architectural performance as limited by DRAM bandwidth. 

Although bandwidths and throughputs can be obtained
from vendor specifications and arithmetic intensities can be
computed by hand, empirical testing more accurately reflects the actual
performance. Acquiring these metrics requires a variety  of
performance profiling tools on the different architectures and machines. For
gathering the bandwidths and throughputs on GPUs, we used \gpumembench
\cite{konstantinidis_quantitative_2017} for measuring the L1 bandwidth and the
Empirical Roofline Tool (Version 1.1.0) \cite{lo_roofline_toolkit_2015}
for measuring all other bandwidths and the peak throughput.  For computing
arithmetic intensities on GPUs, we used NVIDIA's \nvprof (CUDA Toolkit
9.2.88 on MSU HPCC, 9.2.148 on SDSC Comet) to measure memory usage to
calculate arithmetic intensities and total FLOP count to estimate FLOP per
finite volume cell update. To measure memory usage of the different caches, we
specifically measured total memory transactions from global memory to the SMs
(\texttt{gld\_transactions} and \texttt{gst\_transactions}, as a rough proxy
for L1 usage), transactions to and from L2 cache
(\texttt{l2\_read\_transactions} and \texttt{l2\_write\_transactions}), and
transactions to and from DRAM/HBM (\texttt{dram\_read\_transactions} and
\texttt{dram\_write\_transactions}). Since we do not use atomic memory
operations, texture memory, or shared memory, we measured zero transactions
from these memory spaces.  For Intel CPUs and KNLs, we used Intel Advisor's
(version 2019 update 5) built-in hierarchical roofline gathering tools to
collect memory bandwidths, throughputs, and arithmetic intensities
\cite{intel_advisor_roofline} using the arithmetic intensity from the
cache-aware roofline model for the roofline of the highest memory level.  For
both CPUs and GPUs, we use total memory transactions to cores and SMs as a
surrogate for L1 cache usage due to limitations in the memory transaction
metrics available. Although some of the memory transactions may not be through
L1 cache, in a best case performance scenario the memory transactions to the
registers are limited by the fastest cache bandwidth, which is the L1 cache
bandwidth.

We used a 3D linear wave on a $256^3$ cell grid for benchmarking \kathena's
performance and arithmetic intensities for the roofline model. Our metric for CPU
machines are for two sockets on a node while the metric for KNLs and GPUs are for
a single device, or a single GK210 chip for the Tesla K80.  In all cases we
found that \kathena's performance is limited by the main memory space that
accommodates the data for a single \mpi task. For GPUs, this is on device
DRAM/HBM, for CPUs this is the DDR3/DDR4 DRAM, and for KNLs this was the
MCDRAM. This result is expected, since the finite volume MHD method in \kathena
is implemented as a series of simple triple or quadruple for-loop kernels that
loop over the data in a task without explicitly caching data.  Since the data
can only fit in its entirety in DRAM, it must be loaded from and written to
DRAM within each kernel.  Future improvements can be made to \kathena to
explicitly cache data in smaller 1D arrays and kept in higher level caches.
This would raise the DRAM arithmetic intensity and facilitate faster
throughput \cite{glines_scalable_2015}.  Similar improvements have already
been implemented upstream in \athena. A more complete solution would involve
fusing consecutive kernels into one kernel to reduce DRAM accesses.
Given the virtually identical performance between \athena and \kathena on
CPUs (cf.~\ref{sec:single-scaling}) we expect the roofline model of \athena
to be practically indistinguishable from \kathena on non-GPU platforms.

\subsubsection{Performance portability metric}
\label{sec:performance-portability-metric}

Performance portability is at present nebulously defined. It is generally held
that a performance portable application can execute wide variety of
architectures and achieve acceptable performance, preferably maintaining a
single code base for all architectures. In order to make valid comparisons
between codes, an objective metric of performance portability is needed.

The metric proposed by \cite{Pennycook2019} quantifies performance portability
by the harmonic sum of the performance achieved on each platform, so that
\begin{equation}
  P(a,p,H) = 
  \left\{\begin{matrix}
    \dfrac{|H|}{ \sum_{i \in H} \frac{1}{e(a,p,i)}} & 
        \text{if } i \text{ is supported } \forall i \in H \\
    0 & \text{otherwise}
  \end{matrix}\right.
\end{equation}
where $H$ is the space of all relevant platforms and $e(a,p,i)$ is the performance
efficiency of application $a$ to solve the problem $p$ on a platform $i$. If an
application does not support a platform, then it is not performance portable
across the platforms and is assigned a metric of 0. The performance
efficiency can also be defined as either the application efficiency, the
fraction of the performance of the fastest application that can solve the
problem on the platform; or as the  architectural efficiency, the achieved
fraction of the theoretical peak performance limited by the hardware that we
computed in Sec. \ref{sec:roofline-models}. Since we did not have MHD codes
implementing the same method as \kathena on all architectures, we used the
architectural efficiencies obtained from the roofline model to compute the
performance portability metric.  For completeness, we considered the
architectural efficiencies as limited by the both the L1 cache and DRAM
bandwidths to compute separate performance portability metric against both
memory spaces.

\begin{figure}[htbp]
\centering
\includegraphics[width=3.5in]{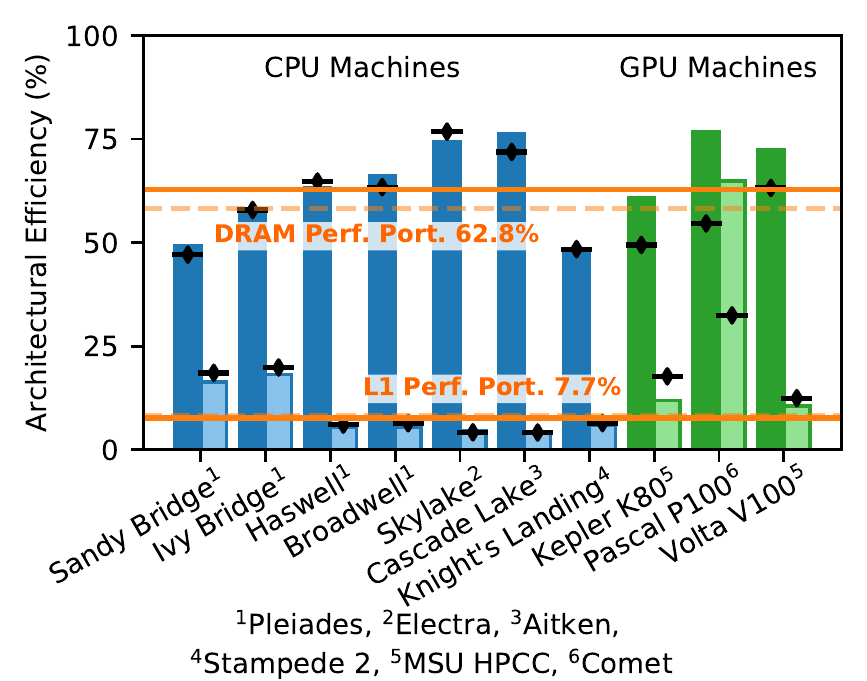}
\caption{Performance Portability plot of several CPU and GPU machines with
  different architectures. Individual bars show the performance of \kathena
  compared to the theoretical peak performance limited by the empirically
  measured DRAM and L1 bandwidths. Black bars with diamonds denote the
  theoretical performance limited by the manufacturer reported bandwidths. The
  performance portability metrics across all architectures for DRAM and L1 are
  shown with horizontal orange lines where solid orange used the empirically
  measured bandwidths and dashed orange uses manufacturer reported bandwidths.\protect\footnotemark
  }
\label{fig:performance-portability}
\end{figure}
\footnotetext{The high L1 efficiency on the NVIDIA Tesla Pascal P100 is
due to a lower obtainable bytes loaded to L1 per cycle compared to the Kepler
and Volta GPUs \cite{jia_dissecting_2018,jia_dissecting_2019}. The lower
L1 cache performance makes it easier to obtain a higher efficiency.}

In Fig.~\ref{fig:performance-portability}, the architectural efficiencies
as measured against the DRAM bandwidth and L1 cache bandwidth are shown with
the computed performance portability metrics.
\kathena achieved $62.8\%$ DRAM
performance portability and $7.7\%$ L1 cache performance portability,
measured across a number of CPU and GPU architectures. In general, \kathena
achieved higher efficiencies on newer architectures.

\subsection{Scaling}

\subsubsection{Single CPU and GPU performance}
\label{sec:single-scaling}
\begin{figure}[htbp]
\centering
\includegraphics[width=0.5\textwidth]{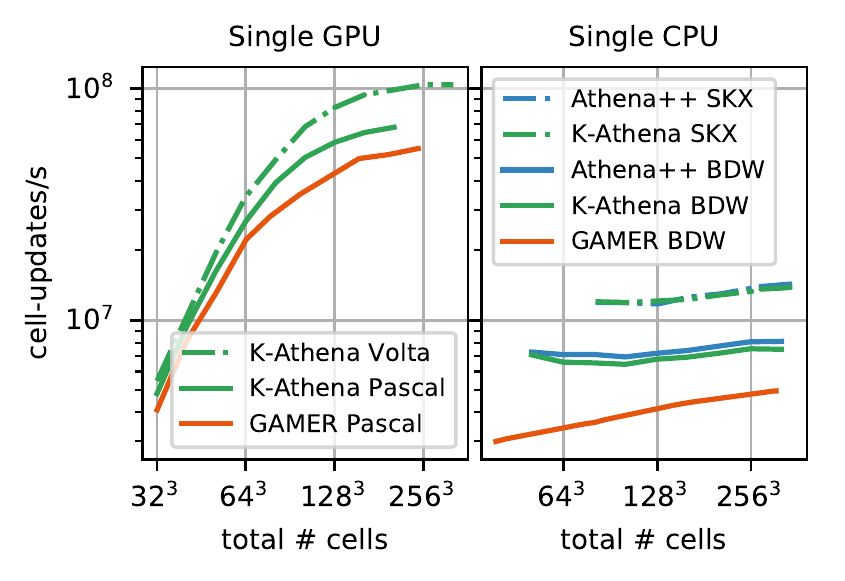}
\caption{Raw performance for double precision MHD (algorithm described in 
Sec.~\ref{sec:results}) 
of \kathena, \athena, and \gamer on a single GPU (left) or CPU (right) for varying problem 
sizes.
Volta refers to an NVIDIA V100 GPU, Pascal refers to an NVIDIA P100 GPU, 
BDW (Broadwell) refers to a 14-core Xeon E5-2680 CPU,  and
SKX (Skylake) refers to a 20-core Xeon Gold 6148 CPU.
The \gamer numbers were reported in \cite{Zhang2018} for the same algorithm used here.
}
\label{fig:single-node}
\end{figure}
In order to compare the degree to which the refactoring of \athena affected performance
we first compare \athena and \kathena on a single CPU.
The right panel of Fig.~\ref{fig:single-node} shows the cell-updates/s achieved 
on an Intel Broadwell and an Intel Skylake CPU for both codes for varying problem size.
Overall, the achieved cell-updates/s are practically independent of problem 
sizes reaching $\approx8\times10^6$ on a single Broadwell CPU and 
$\approx1.4\times10^7$ on a single Skylake CPU.
Moreover, without any additional performance optimizations (see discussion 
in Sec.~\ref{sec:limit-and-enhance}), \kathena is virtually on par with \athena, 
reaching $93\%$ or more of the original performance.
For comparison, we also show the results of \gamer \cite{Zhang2018}.
It is another recent (astrophysical) MHD code with support for CPU and (\cuda-based) 
GPU accelerated calculations and has directly been compared to \athena in \cite{Zhang2018}.
We also find that \athena (and thus \kathena) is about 1.5 times faster than 
\gamer on the same CPU.

A slightly smaller difference (factor of $\approx1.25$)  is observed when comparing 
results for GPU runs
as shown in the left panel of Fig.~\ref{fig:single-node}.
On a P100 Pascal GPU, \kathena is about 1.3 times faster than \gamer, suggesting
that the difference in performance is related to the fundamental code design and not
related to the implementation of specific computing kernels.
On a single V100 Volta GPU, \kathena reaches a peak performance of
greater than $10^8$ cell-updates/s
for large problem sizes.
In general, the achieved performance in cell-updates/s is strongly dependent on the
problem size.
For small grids the performance is more than one order of magnitude lower than what is
achieved for the largest permissible grid sizes that still fit into GPU memory.
The plateau in performance on GPUs at larger grid sizes is due to DRAM
bandwidth  impeding \kathena's performance, as discussed in
Section~\ref{sec:roofline-models}.

\subsubsection{Weak scaling}
\label{sec:weak-scaling}
\begin{figure*}[htbp]
\centering
\includegraphics[width=\textwidth]{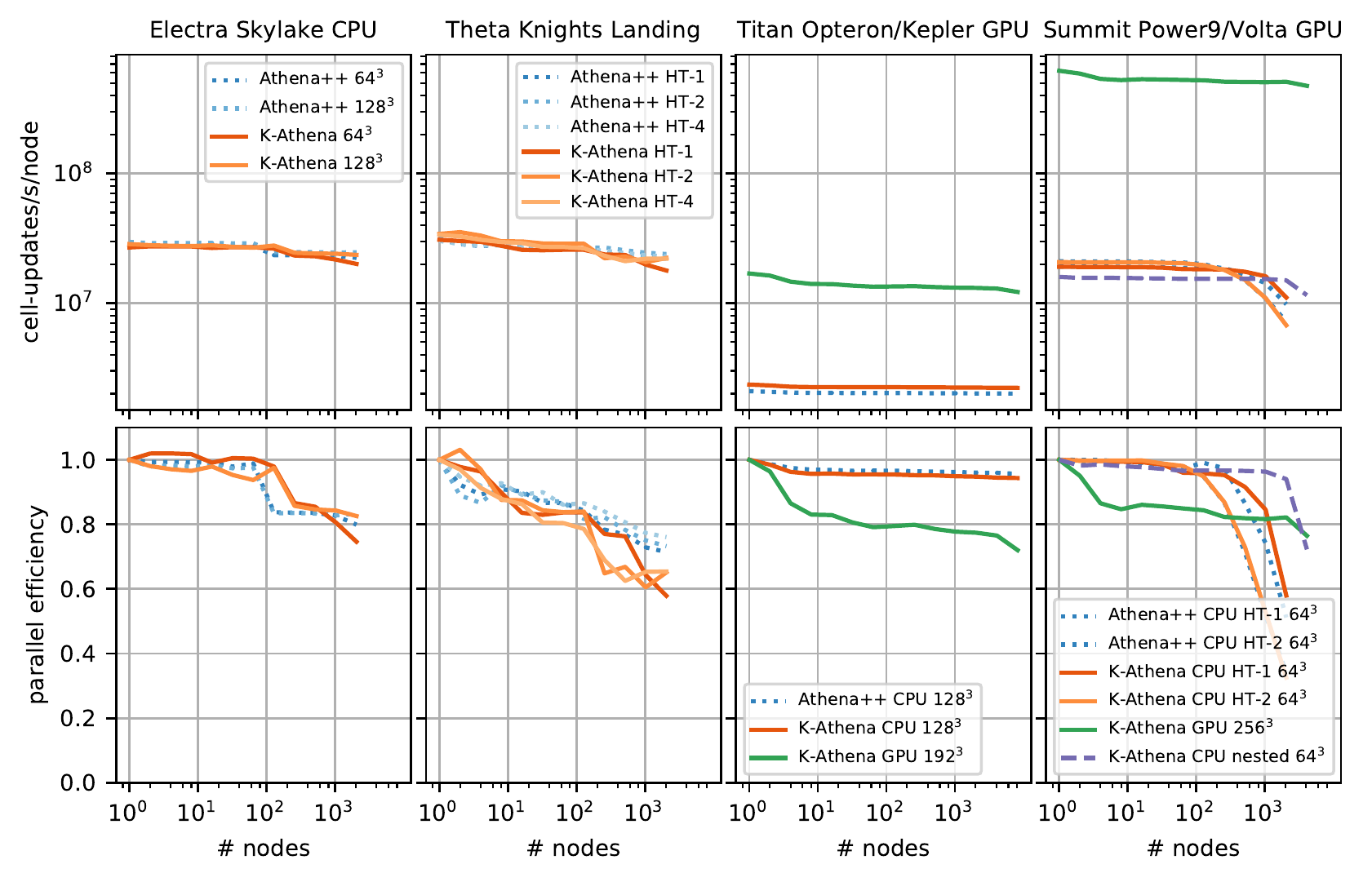}
\caption{
Weak scaling for double precision MHD (exact algorithm described in Sec.~\ref{sec:results}) 
on different supercomputers and 
architectures for \kathena and the original \athena version.
Numbers correspond to the 80th percentile of individual cycle performances of several runs
in order to reduce effects of network variability.
The top row shows the raw performance in number of cell-updates per second per node and 
can directly be compared between different system and architectures.
The bottom row shows the parallel efficiency normalized to the individual single 
node performance.
The first column contains results for a workload of $64^3$ and $128^3$ cells per core
on NASA's Electra system using two 20-core Intel Xeon Gold 6148 processors per node.
The second column shows results for a workload of $64^3$ per core on ALCF's Theta system
with one 64-core Intel Xeon Phi 7230 (Knights Landing) per node.
HT-1, HT-2, and HT-4 refers to using 1, 2, and 4 hyperthreads per core, respectively.
The third column shows results for a workload of $128^3$ per CPU core and $192^3$ per GPU
on OLCF's Titan system with one AMD Opteron 6274 16-core CPU and one NVIDIA K20X (Kepler) 
GPU per node.
The last column contains results for a workload of $64^3$ per CPU core and $256^3$ per GPU
on OLCF's Summit system with two 21-core IBM POWER9 CPUs and six NVIDIA V100 (Volta) GPUs
per node.
On all systems the GPU runs used \texttt{1D} loops and the CPU 
runs used \texttt{simd-for} loops
with the the exception of the dashed purple line on Summit that used \kokkos nested 
parallelism, see Sec.~\ref{sec:kathena-loops} for more details.
}
\label{fig:weak-scaling}
\end{figure*}
Weak scaling results (using the same test problem and algorithm as in Sec.~\ref{sec:single-scaling}) 
for \kathena and the original \athena version on 
different systems and architectures are shown in Fig.~\ref{fig:weak-scaling}.
Note that the chosen problem setup (using a single meshblock per MPI process)
is effectively not making use of 
of the asynchronous communication capabilities to allow for overlapping
computation and communication.

Overall, the differences between \kathena and \athena on CPUs and Xeon Phis are marginal.
This is expected as \kathena employed \texttt{simd-for} loops for all
kernels that are  
similar to the ones already in \athena.
Therefore, the parallel efficiency is also almost identical between both codes, reaching
$\approx80\%$ on NASA's Electra system with Skylake CPUs 
(first column in Fig.~\ref{fig:weak-scaling})
and $\approx70\%$ on  ALCF's Theta system with Knights Landing Xeon Phis 
(second column in Fig.~\ref{fig:weak-scaling})
at 2,048 nodes each.
Using multiple hyperthreads per core on Theta has no significant influence on the results 
given the intrinsic variations observed on that system\footnote{According to the ALCF
support staff, system variability contributes around 10\% to the
fluctuations in performance between identical
runs.}.

The first major difference is observed on OLCF's Titan 
(third column in Fig.~\ref{fig:weak-scaling}),
where results for \kathena on GPUs are included.
While the parallel efficiency for both codes remains at 94\% up to 8,192 
nodes using only CPUs, it drops to 72\% when using GPUs with \kathena.
However, the majority of loss in parallel efficiency already occurs going from 1 to 8 nodes
using GPUs and afterwards remains almost flat.
This behavior is equally present for CPU runs but less visible due to the higher
parallel efficiency in general.
The differences in parallel efficiency between CPU and GPU runs can be attributed to the
vastly different raw performance of each architecture.
On a single node the single Kepler K20X GPU is about 7 times faster than
the 16-core AMD Opteron CPU.
Given that the interconnect is identical for GPU and CPU communication, the effective ratio
of computation to communication is worse for GPUs.
Despite the worse parallel efficiency on GPUs the raw per-node performance using GPUs
is still about 5.5 times faster than using CPUs at 8,192 nodes,
which is overall comparable to the ratio of theoretical peak performances in both
FLOPS and DRAM bandwidth.

\kathena on 
OLCF's Summit system (last column in Fig.~\ref{fig:weak-scaling}) with six Volta V100 GPUs
and two 21-core POWER9 CPUs exhibits a GPU weak scaling behavior similar to the one observed
on Titan.
Going from 1 to 8 nodes results in a loss of 15\% and afterwards the parallel
efficiency remains almost flat to 76\% on 4,096 nodes.
The CPU weak scaling results for both codes using CPUs reveal properties of the 
interconnect.
The weak scaling is almost perfect up to 256 nodes using 1 hyperthread per core 
and afterwards rapidly plummets.
Using 2 hyperthreads per core (i.e., doubling the number of threads making \mpi calls and 
doubling the number of \mpi messages sent and received, as described in 
Sec.~\ref{sec:athena})
the steep drop in parallel efficiency is already observed 
beyond 128 nodes.
No such drop is observed using GPUs, which perform $42/6 = 7$ times fewer \mpi calls 
(compared to using 1 hyperthread per core) with larger message sizes in general.

Naturally, this is tightly related to the existing communication pattern in \athena, 
i.e., coarse grained threading over meshblocks with each thread performing one-sided
\mpi calls. 
Without making additional changes to the code base, we can evaluate the effect of
reducing the number of \mpi calls for a fixed problem size in a multithreaded CPU setup 
using \kokkos nested parallelism in \kathena.
More specifically, we use the triple nested construct illustrated in Listing~\ref{lst:TeamPolicy}
allowing multiple threads handling a single meshblock.
As a proof of concept, the results for using using 1 \mpi process per 2 cores each with 
one thread are shown in the purple dash line 
in the last column of Fig.~\ref{fig:weak-scaling}.
While the raw performance on a single node is slightly lower (about 16\%), the improved 
communication pattern results in a higher overall performance for $>1{,}024$ nodes.
Similarly, the sharp drop in parallel efficiency has been shifted to first occur at
2,048 nodes.

At the single node level the six GPUs on Summit are tightly connected via NVLink.
The weak scaling efficiency from one GPU to six GPUs on a single node
is $\approx99\%$ (cf., $>6\times10^8$ cell-updates/s/node for a single node in the 
top right panel of Fig.~\ref{fig:weak-scaling}).
In addition, the host interconnect has a lower bandwidth and higher latency compared to NVLink.
Thus, the intra-node parallel overhead is generally negligible in our analysis.

Finally, the raw per-node performance is overall comparable between Intel Skylake CPUs, 
Intel Knight Landing Xeon Phis, IBM POWER9 CPUs, and a single NVIDIA Kepler GPU, ranging between 
$\approx$\,1.5 -- 3$\times10^7$ cell-updates/s/node.
The latest NVIDIA Volta GPU is a notable exception, reaching more than
 $10^8$ cell-updates/s/GPU.
This performance, in combination with six GPUs per node on Summit and a high parallel 
efficiency, results in a total performance of $1.94\times10^{12}$ cell-updates/s on
4,096 nodes.

\subsection{Strong scaling}
\begin{figure}[htbp]
\centering
\includegraphics[width=0.5\textwidth]{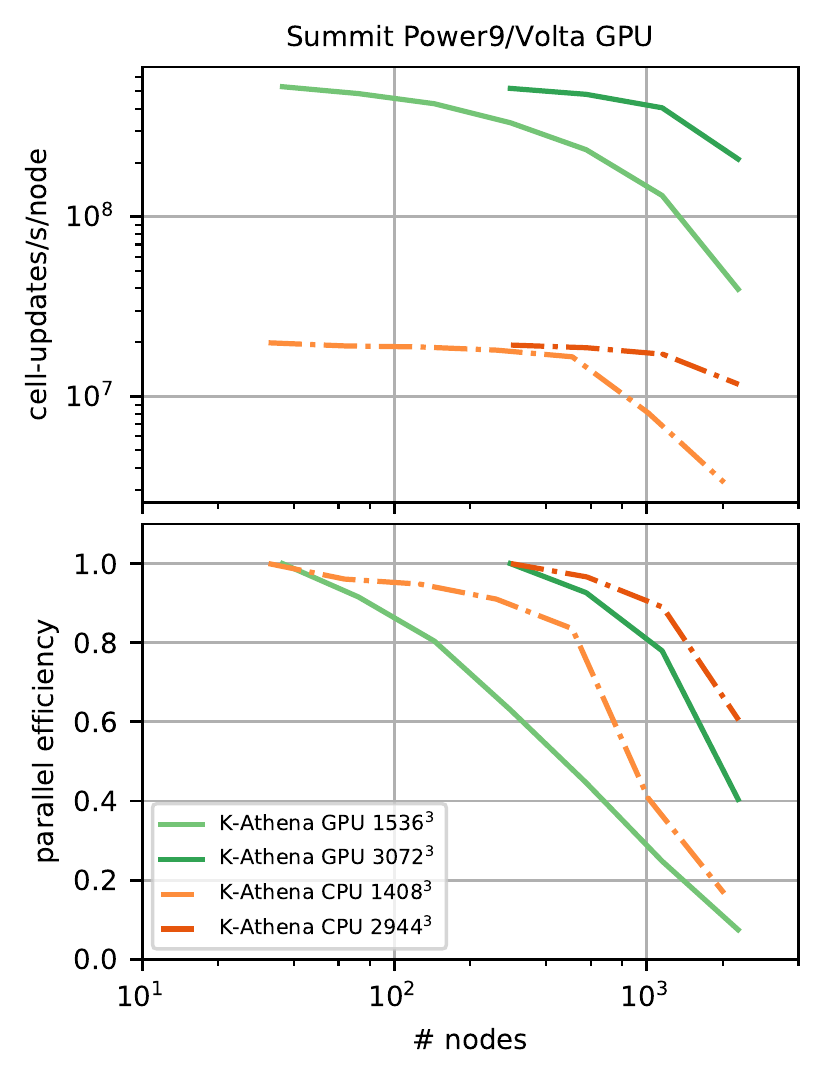}
\caption{Strong parallel scaling for double precision MHD (algorithm described in 
Sec.~\ref{sec:results}) 
of \kathena on NVIDIA V100 GPUs (6 GPUs per node; green solid lines) 
and IBM Power 9 CPUs (42 cores per node; orange/red dash dotted lines)
on Summit.
The top panel shows the raw performance in cell-updates per second per node and
the bottom panel shows the parallel efficiency.
The effective workload per GPU goes from $256^3$ to $64^3$ for the $1{,}536^3$ domain and
from $256^3$ to $128^3$ for the $3072^3$ domain.
In the CPU case the effective workload per single Power9 CPU (21 cores) goes from
$353^3$ to $88^3$ for the $1{,}408^3$ domain and from $353^3$ to $177^3$ for the $2{,}944^3$
domain.
The resulting effective workloads per node are comparable (within few percent) between 
GPU and CPU runs.
}
\label{fig:strong-scaling}
\end{figure}

Strong scaling results for \kathena on Summit on both CPUs and GPUs are shown in
Fig.~\ref{fig:strong-scaling} (same test problem and algorithm as in
Sec.~\ref{sec:single-scaling}).
Overall, strong scaling in terms of parallel efficiency is better on CPUs than on GPUs.
For example, for a $1{,}408^3$ domain the parallel efficiency using CPUs remains $>83\%$ 
going from 32 to 512 nodes whereas it drops to $45\%$ for the similar GPU case ($1{,}536^3$ 
domain using 36 to 576 nodes).
This is easily explained by comparing to the single CPU/GPU performance discussed in
Sec.~\ref{sec:single-scaling}, which effectively corresponds to on-node strong scaling.
The more pronounced decrease in parallel efficiency on the GPUs is a direct result of
the decreased raw performance of GPUs with smaller problem sizes per GPU.
The increased communication overhead of the strong scaling test plays only a secondary 
role.
Therefore, the strong scaling efficiency of \kathena in comparison to
\athena is expected to be identical.
Moreover, additional performance improvements, as discussed in the following Section,
will greatly benefit the strong scaling behavior of GPUs in general.
Nevertheless, the raw performance of the GPUs still outperforms CPUs
by a large multiple
despite the worse strong scaling parallel efficiency.
For example, in the case discussed above on Summit, the per-node performance of GPUs over CPUs
is still about 14 times higher at $>512$ nodes.

\section{Current limitations and future enhancements}
\label{sec:limit-and-enhance}
Our primary goal for the current version of \kathena was to make GPU-accelerated 
simulations possible while maintaining CPU performance, and to do so with the smallest amount of
code changes necessary.
Naturally, this resulted in several trade-offs and leaves room for further (performance)
improvements in the future.

For example, we are currently not making use of the memory hierarchy abstraction
provided by \kokkos.
This includes  more advanced hardware features such as
scratch spaces on GPUs.
Scratch space can be shared among threads of a \texttt{TeamPolicy} and allows for efficient
reuse of memory.
We could use scratch space to reduce the number of reads from DRAM in stenciled
kernels (like the fluid solver's reconstruction step). We could also fuse consecutive kernels
to further reduce reads and writes to DRAM, although this would also increase
register and possibly spill store usage.
Moreover, complex kernels such as a Riemann solver could be broken down further by using
\texttt{TeamThreadRange}s and \texttt{ThreadVectorRange}s structures that are closer to
the structure of the algorithm.
This is in contrast to our current approach where all kernels are treated equally, with the 
same execution policies independent of the individual algorithms within the kernels.
The Riemann solver could also be split into separate kernels to reduce the
number of registers needed, eliminate the use of spill stores on the GPU, and
allow higher occupancy on the GPU.

Similarly, on CPUs and Xeon Phis we are currently not using a \kokkos parallel execution 
pattern.
The macro we introduced to easily exchange parallel patterns replaces the parallel region
on CPUs and Xeon Phis with a simple nested \texttt{for} loop including a \texttt{simd} 
pragma, as shown in Listing~\ref{lst:simd-for}.
This is required for maximum performance 
as the implicit \verb=#pragma ivdep= hidden in the \kokkos templates is less
aggressive than the explicit  \verb=#pragma omp simd= with respect to automated
vectorization.
We reported this issue and future \kokkos updates will address this by either providing an
explicit tightly nested vectorized loop pattern and/or adding support for a \verb=simd= 
property to the execution policy template.

Another possible future improvement is an increase in parallel efficiency by overlapping
communication and computation.
While \athena is already built for asynchronous communication 
through one-sided \mpi calls
and a task based execution
model, more fine-grained optimizations are possible.
For example, spatial dimensions in the variable reconstruction step that occurs after 
the exchange of 
boundary information could be split, so that the kernel in the first dimension could run
while the boundary information of the second and third dimension are still being exchanged.
In addition, the next major \kokkos release will contain more support
for architecture-dependent task based execution and, for example, will allow for the transparent use of 
\cuda streams.

\cuda streams may also help in addressing another current limitation of \kathena on GPUs.
Our minimal implementation approach currently limits all \texttt{meshblock}s to be allocated
in a fixed memory space.
This means that the total problem size that can currently be addressed with \kathena is 
limited by the total amount of GPU memory available.
An alternative approach is keeping the entire mesh in system memory, which is still several
times larger than the GPU memory on most (if not all) current machines.
For the execution of kernels individual \texttt{meshblock}s would be copied back and forth
between system memory and GPU memory.
Here, \cuda streams could be used to hide these expensive memory transfers as they would occur
in the background while the GPU is executing different kernels.
Theoretically, meshes larger than the GPU memory could already be used right now with the
help of unified memory.
However, given that the code is not optimized for efficient page migrations the resulting
performance degradation is large (more than a factor of 10). 
Thus, using unified memory with meshes larger than the GPU memory is not recommended.

\section{Conclusions}
\label{sec:conclusions}

We presented \kathena~-- a \kokkos-based performance portable version of the finite 
volume MHD code \athena.
\kokkos is a C++ template library that provides abstractions for on-node
parallel regions and the memory hierarchy.
Our main goal was to enable GPU-accelerated simulations while maintaining \athena's
excellent CPU performance using a single code base and with minimal changes to the
existing code.

Generally, four main changes were required in the refactoring process.
We changed the underlying memory management in \athena's multi-dimensional array class
to make transparently use of \kokkos's equivalent multi-dimensional arrays, i.e.,
\verb=Kokkos::View=s.
We exchanged all (tightly) nested \verb=for= loops with the \kokkos equivalent
parallel region, e.g., a \verb=Kokkos::parallel_for=, which are now kernels that
can be launched on any supported device.
We inlined all support functions (e.g., the equation of state) that are called within 
kernels.
We changed the communication buffers to be \verb=View=s so that \mpi calls between GPUs
buffers are directly possible without going through system memory.

With all changes in place we performed both profiling and scaling studies 
across different platforms, including NASA's Electra system with Intel Skylake CPUs,
ALCF's Theta system with Intel Xeon Phi Knights Landing, OLCF's Titan with AMD 
Opteron CPUs and NVIDIA Kepler GPUs, and OLCF's Summit machine with IBM Power9 CPUs
and NVIDIA Volta GPUs.
Using a roofline model analysis, we demonstrated that the current implementation of the
MHD algorithms is memory bound by either the DRAM, HBM, or MCDRAM bandwidths on CPUs and GPUs.
Moreover, we calculated a performance portability metric of 62.8\% across Xeon Phis, and
6 CPU and 3 GPU generations.

Detailed \kokkos profiling revealed that there is currently no universal \kokkos
execution policy (how a parallel region is executed) 
that achieves optimal performance across different architectures.
For example, a one-dimensional loop with manual index matching from 1 to 3D/4D is
fastest on GPUs (achieving $>10^8)$ double precision MHD cell-updates/s on a single 
NVIDIA V100 GPU) whereas tightly nested \verb=for= loops with \verb=simd= directives
are fastest on CPUs.
This is primarily a result of \kokkos's specific implementation details
and expected to improve in future releases through more flexible execution policies.

Strong scaling on GPUs is currently predominately limited by individual GPU
performance and not by communication.
In other words, insufficient GPU utilization outweighs additional performance
overhead with decreasing problem size per GPU.

Weak scaling is generally good, with parallel efficiencies of $80\%$ and higher 
for more than 1,000 nodes across all machines tested.
Notably, on Summit \kathena achieves a total calculation speed of $1.94\times10^{12}$ cell-updates/s
on 24,567 V100 GPUs at a speedup of 30 compared to using the available 172,032 CPU cores.

Finally, there is still a great deal of untapped potential left, e.g., using
more advanced hardware features such as fine-grained nested parallelism, scratch pad 
memory (i.e., fast memory that can be shared among threads), or \cuda streams.
These are currently being addressed within the new \parthenon collaboration
(\url{https://github.com/lanl/parthenon}), which is developing a performance portable
adaptive mesh refinement framework based on the results presented here.

Nevertheless, we achieved our primary performance portability  goal of 
enabling GPU-accelerated simulations while maintaining CPU performance using a single 
code base.
Moreover, we consider the current results highly encouraging and will continue with
further development on the project's GitLab repository at 
\url{https://gitlab.com/pgrete/kathena}.
Contributions of any kind are welcome!

\ifCLASSOPTIONcompsoc
  \section*{Acknowledgments}
\else
  \section*{Acknowledgment}
\fi

The authors would like to thank 
the \kokkos developers, particularly Christian Trott and Steve Bova, and the
organizers of the 2018 Performance Portability with \kokkos Bootcamp for
their help using \kokkos in \athena.
We thank Kevin O'Leary and Dunni Aribuki for support with the Intel Advisor.
We thank Kristian Beckwith for inspiring discussions on \kokkos.
We thank the \athena team for making
their code public and for their well designed code.  
We acknowledge funding by NASA Astrophysics Theory Program grant \#NNX15AP39G.
Code development, testing, and benchmarking was made possible through various
computing grants including allocations on NASA Pleiades (SMD-16-7720), 
OLCF Titan (AST133), OLCF Summit (AST146), ALCF Theta (athena\_performance),
XSEDE Comet (TG-AST090040), 
and Michgian State University's High Performance Computing Center.

\ifCLASSOPTIONcaptionsoff
  \newpage
\fi



\bibliographystyle{IEEEtran}
\bibliography{references}
%\begin{thebibliography}{1}
%\end{thebibliography}


\begin{IEEEbiography}[{\includegraphics[width=1in,height=1.25in,clip,keepaspectratio]{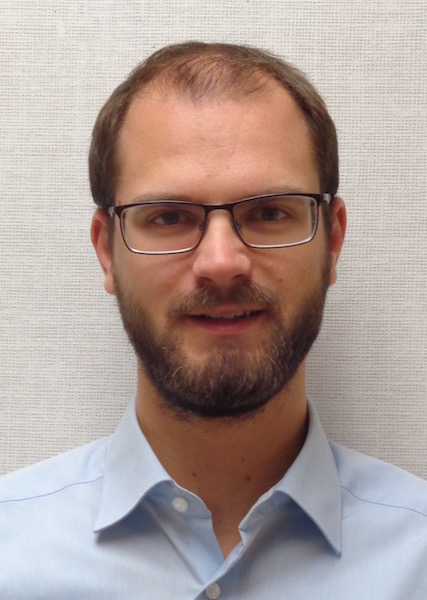}}]{Philipp Grete}
After receiving a B.Sc. in Computer Science in 2008 from the University
of Cooperative Education in Stuttgart, Germany, Philipp Grete worked for
Hewlett-Packard before studying Physics (B.Sc.) and Computer Science
(M.Sc.) at the University of Göttingen, Germany, from 2010 to 2013.
In his Ph.D. thesis (2014-2016, University of Göttingen, Germany) he
worked on subgrid-scale modeling of compressible magnetohydrodynamic
turbulence. Since October 2016, he is a postdoctoral research associate
at Michigan State University.
His current research interest include fundamental processes 
in magnetized astrophysical fluids, numerical methods in computational fluid dynamics, and 
high-performance computing with an emphasis on performance portability.
\end{IEEEbiography}

\begin{IEEEbiography}[{\includegraphics[width=1in,height=1.25in,clip,keepaspectratio]{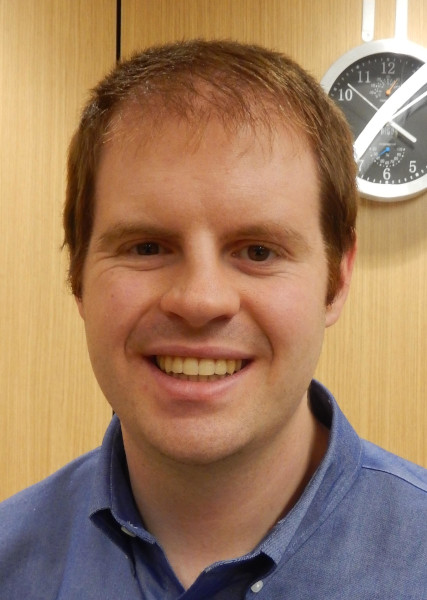}}]{Forrest~W.~Glines}
Forrest Glines is PhD student at Michigan State University working towards a
  dual degree in Astrophysics and Computational Mathematics, Science, and
  Engineering. He received his B.S.~in Physics from Brigham Young University in
  2016, publishing work on  magnetohydrodynamics methods using GPUs. In the
  2015 and 2016 he also spent 6 months collectively at Los Alamos National
  Laboratory, simulating galaxy cluster and developing simulations coupling
  radiative transfer and hydrodynamics.  Forrest's current research interests
  includes the coupling of galaxy clusters with active galactic nuclei, plasma
  turbulence, the evolution of galaxies with magnetic fields, and developing
  simulations for the exascale era.
\end{IEEEbiography}

%\newpage

\begin{IEEEbiography}[{\includegraphics[width=1in,height=1.25in,clip,keepaspectratio]{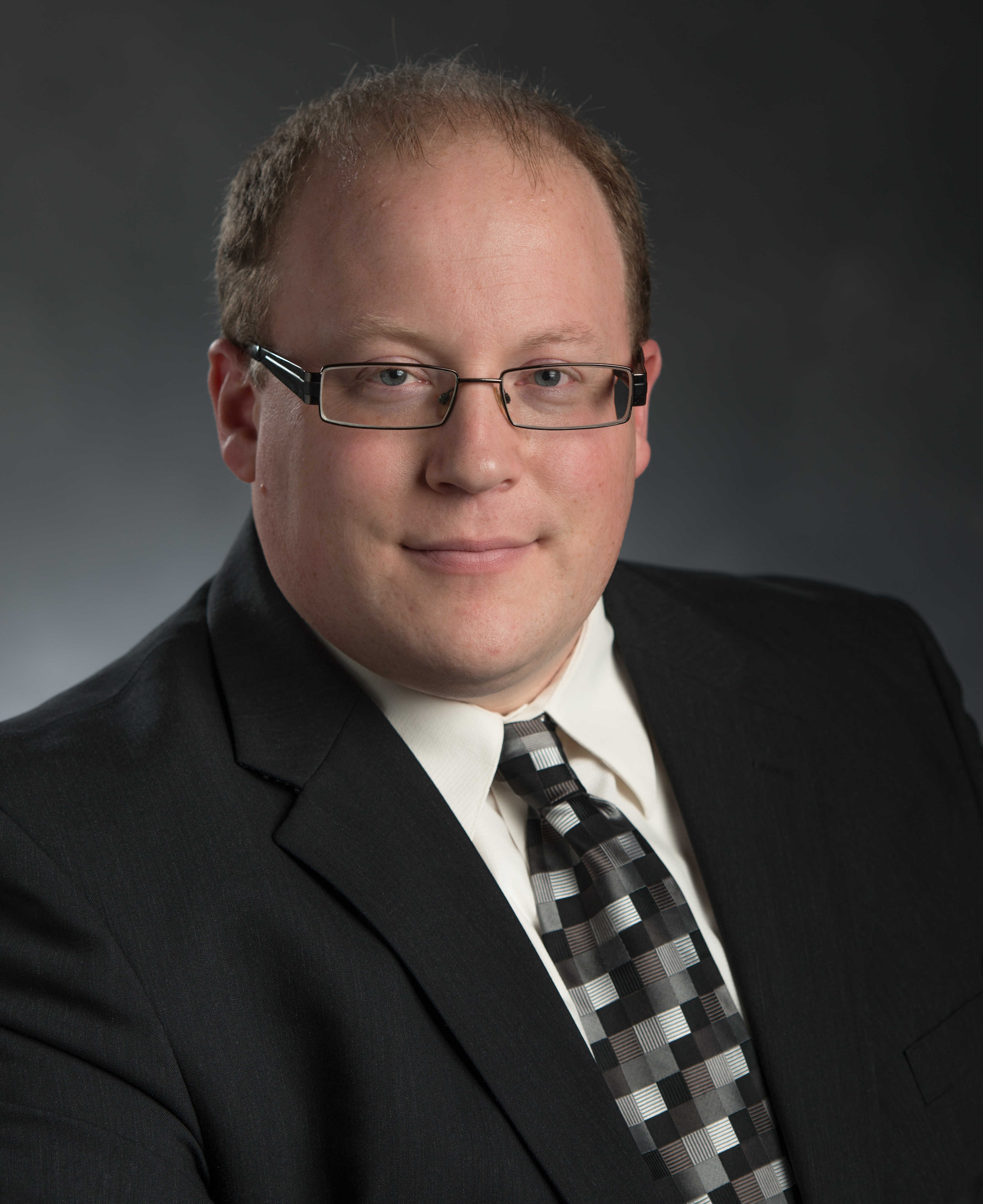}}]{Brian~W.~O'Shea}
Brian O'Shea received his PhD in Physics from the University of Illinois at 
Urbana-Champaign in 2005.
His dissertation research focused on simulations of cosmological structure formation, 
particularly in the first generation of stars in the Universe.  
He was then a Director’s Postdoctoral Fellow at Los Alamos National Laboratory in both the 
Theoretical Astrophysics Group and the Applied Physics Division, where he studied the 
formation of galaxies over a wide range of mass scales and cosmic epochs. 
Since 2008 he has been a professor at Michigan State University with appointments in 
the Department of Computational Mathematics, Science, and Engineering, Department of Physics 
and Astronomy, and National Superconducting Cyclotron Laboratory.  
His current research interests range across a wide variety of topics in cosmological 
structure formation, astrophysical plasma turbulence, high-performance computing, 
and computational science education.  
He is one of the core developers and leaders of the Enzo community code 
(\url{enzo-project.org}), and a contributor to several other open-source projects.
\end{IEEEbiography}


%\vfill


\end{document}